\documentstyle[aps,floats,eqsecnum,psfig]{revtex}

\begin{document}
\draft

\title{Viscoelastic effects in a spherical Gravitational Wave antenna}

\author{J.A. Lobo {\it and\/} J.A. Ortega}
\address{Departament de F\'\i sica Fonamental \\
         Universitat de Barcelona, Spain.}

\date{\today}
\date{19 Jabuary 1998) \\ (Revised \today}

\maketitle

\begin{abstract}

Internal friction effects are responsible for line widening of the
resonance frequencies in spherical gravitational wave detectors, and result
in exponentially damped oscillations of its eigenmodes with a decay time
of the order of $Q/\omega\/$. We study the solutions to the equations of
motion for a viscoelastic spherical GW detector based on various different
assumptions about the material's constituent equations. Quality factor
dependence on mode frequency is determined in each case, and a discussion
of its applicability to actual detectors is made.

\end{abstract}
% \pacs{04.80.Nn, 95.55.Ym}
\pacs{{\sc Key words}: Spherical GW detector, viscosity, linewidth}

\section{Introduction}

Spherical Gravitational Wave (GW) detectors will almost certainly be the
next generation of resonant antennae, due to their multimode-multifrequency
capabilities \cite{clo,ls,jm,nadja}, as well as their potentially enhanced
sensitivity relative to their currently operating cylindrical counterparts
\cite{nau,as,alle,niobe}. Conviction that this is going to be the case has
encouraged a remarkable research effort within the GW community, and a
variety of important topics has been addressed, ranging from theoretical to
practical aspects of the problem. Several countries worldwide are currently
developing projects to build and operate large mass ($\sim$100 tons) spherical
detectors \cite{omni}.

Sensitivity analyses of the response of a solid elastic sphere to an
incoming GW excitation are usually made under the idealised assumption that
there are no dissipative effects in the antenna \cite{lobo,wp,bian,vega}. In
actual practice, however, such effects do exist, and translate into finite
{\it decay times\/} of the solid's oscillations or, equivalently, into
finite {\it linewidths\/} in the system Fourier spectrum. Those decay times
are generally much longer than the inverse frequencies of the
eigenmodes~\cite{cff}, so spectral lines are actually very narrow, and peak
at the values predicted by the ideal (frictionless) equations of Elasticity
\cite{phd}. It thus appears appropriate to model the system behaviour
as undergoing {\it linear\/} dissipative effects, so that its eigenmode
amplitudes decay {\it exponentially}, i.e., they are proportional to the
expression

\begin{equation}
        e^{-\omega t/Q}\,\sin\omega t   \label{I.1}
\end{equation}
rather than to the non-dissipative $\sin\omega t$. Here the {\it quality
factor Q\/} is assumed constant (i.e., time independent), and accounts for
the linewidth of the mode with frequency $\omega\/$. Although it is well
known \cite{cff} that this quality factor is different for different
oscillation modes, proper understanding of its variablilty is not quite as
satisfactory as other aspects of the physics of the spherical GW antenna.
This is partly due to the difficulties to set up a model to adequately
describe the internal friction within an elastic solid body.

In the specialised literature on the subject (see \cite{ja} and references
therein), viscoelastic effects are often described by means of so called
{\it constituent equations\/}. These are extensions of Hooke's law relating
the stress and strain tensors in the solid. Much like in a simple
one-dimensional spring, internal friction forces can be considered
proportional to the instantaneous velocity of the oscillating mass, hence
constituent equations usually contain time derivatives of those tensors, and
depend on a small number of viscosity parameters, to be added to the elastic
Lam\'e coefficients $\lambda$ and $\mu\/$ ---see below. There is however no
unique way in which this idea can be carried over from a one-dimensional
system into a set constituent equations for a three-dimensional solid, and
so different alternatives result in different models for the purpose.

In this paper we propose to study and discuss the results of applying to a
spherical GW detector the equations of various such phenomenological models,
in view to determine how quality factors change from mode to mode in each
case. This, it is hoped, will generate some insight into the nature of the
viscous processes which take place in a specific spherical GW detector, and
help to assess on the basis of spectral measurements which particular class
of viscoelastic solid a given material belongs to. In turn, better
understanding of material's macroscopic properties should also contribute
relevant information to the currently important issue of spherical GW
detector design \cite{sfera,grail}. In section 2 we present the general
equations, and in the subsequent sections we successively consider the
Kelvin-Voigt, Maxwell, Standard Linear and Genaralised Mechanical models.
As we shall see, the sphere's vibration eigenmodes always group into the
usual families of {\it toroidal\/} (purely torsional) and {\it spheroidal\/}
modes, but different quality factor dependences arise in different models.
In section 6 we present a summary of conclusions, and two appendices are
added to clarify a few mathematical technicalities.

\section{The general equations}

GWs bathing the Earth are known to be extremely weak ---see e.g. \cite{sh}
for a review---, so that the classical equations of linear Elasticity are
very good to describe the GW induced motions of a spherical antenna in
the expected frequency range, roughly 10$^2$--10$^4$ Hz. These equations
are \cite{ll}

\begin{equation}
    \rho\,\frac{\partial^2 s_i}{\partial t^2} -
    \frac{\partial\sigma_{ij}}{\partial x_j} = f_i  \label{eqmov}
\end{equation}
where ${\bf s}({\bf x},t)$ is the field of displacements in the elastic
body, and ${\bf f}({\bf x},t)$ is the GW induced density of forces acting
on the solid ---see \cite{lobo} for full technical details. $\sigma_{ij}\/$
is the the {\it stress\/} tensor, and is related to the {\it strain\/}
tensor

\begin{equation}
        s_{ij}\equiv\frac{1}{2}\,\left(\frac{\partial s_i}{\partial x_j}
               + \frac{\partial s_j}{\partial x_i}\right)    \label{I.2}
\end{equation}
through a set of {\it constituent equations\/}. In the case of a
non-dissipative (ideal) solid these are simply the expression of Hooke's
law. In a dissipative one these equations include {\it time derivatives\/}
of both $s_{ij}\/$ and $\sigma_{ij}\/$ to account for internal friction
effects. Constituent equations are of the following general type:

\begin{equation}
   L(s_{ij},\dot s_{ij},\ddot s_{ij},\ldots;
     \sigma_{ij},\dot\sigma_{ij},\ddot\sigma_{ij},\ldots) = 0
   \label{I.3}
\end{equation}

In this paper we shall limit ourselves to {\it linear\/} constituent
equations, an excellent approach for a GW detector, as already stressed.
In the simplest instance, only first order derivatives will appear in
(\ref{I.3}), and we shall consider this first. Then we shall also devote
some attention to more complicted models.

The equations of motion (\ref{eqmov}) must be supplemented with suitable
{\it boundary conditions\/}. We shall prescribe the usual ones

\begin{equation}
  \sigma_{ij}\,n_j = 0 \qquad {\rm at} \qquad r=R     \label{I.4}
\end{equation}
where {\bf n} is a unit outward-pointing vector, expressing that the
surface of the sphere is free from any tensions and/or tractions.

\section{Kelvin--Voigt model}

This model assumes that the solid is homogeneous and isotropic, and is
characterised by the following constituent equations \cite{ja}:

\begin{equation}
  \sigma_{ij}=\left(\lambda+\lambda'\frac{\partial}{\partial t}\right)
   s_{kk}\,\delta_{ij} + 2\left(\mu+\mu'\frac{\partial}{\partial t}\right)
   s_{ij}   \label{consKV}
\end{equation}

The constants $\lambda$ and $\mu$ are the usual Lam\'e coefficients
describing the purely elastic behaviour of the body \cite{ll}, while the
positive coefficients $\lambda'$ and $\mu'$ parametrise its {\it viscous\/}
properties\footnote{
These coefficients are actually analogous to those which describe the
viscosity of fluids in Hydrodynamics: {\it shear\/} viscosity ($\mu'$) and
{\it bulk\/} viscosity ($2\mu'+3\lambda'$).}, which are proportional to the
change rate of the strain tensor, $\partial_t s_{ij}$.

If equations (\ref{consKV}) are replaced into (\ref{eqmov}) we obtain the
equations of motion for ${\bf s}({\bf x},t)$:

\begin{equation}
      \rho\,\frac{\partial^2 {\bf s}}{\partial t^2} =
      \left(\mu + \mu'\frac{\partial}{\partial t}\right)\nabla^2 {\bf s} \,-
      \,\left[(\lambda+\mu)+(\lambda'+\mu')\frac{\partial}{\partial t}\right]
      \,\nabla(\nabla{\bf \cdot}{\bf s}) + {\bf f}({\bf x},t)
      \label{eqmovKV}
\end{equation}

The solution to this system of coupled equations can be expressed in terms
of a Green function, the construction of which requires explicit knowledge
of its {\it eigenmode\/} solutions. As is well known, the latter correspond
to the free oscillations of the solid ---no density of external forces in the
rhs of (\ref{eqmovKV}). As usual, we attempt to find such eigen-solutions
in the factorised form

\begin{equation}
     {\bf s}({\bf x},t)=T(t)\,{\bf s}({\bf x})    \label{separ}
\end{equation}
which results in the following (dots on symbols stand for time derivatives):

\begin{equation}
  \rho\ddot T(t) {\bf s}({\bf x}) = \left[(\lambda+\mu)T(t) +
   (\lambda'+\mu')\dot T(t)\right]
   \nabla\left(\nabla{\bf \cdot}{\bf s}({\bf x})\right) +
   \left[\mu\,T(t)+\mu'\,\dot T(t)\right]
   \nabla^2{\bf s}({\bf x})
   \label{eqmovKV2}
\end{equation}

Next we use the well known \cite{ll} decomposition of a three dimensional
vector field into its irrotational a divergence free components:

\begin{equation}
  {\bf s}({\bf x}) = {\bf s}_l({\bf x}) + {\bf s}_t({\bf x})\ ,
  \qquad \nabla{\bf \cdot}{\bf s}_t({\bf x}) =
  \nabla\times{\bf s}_l({\bf x}) = 0      \label{decom}
\end{equation}

By the methods described in Appendix A it can be seen that the equations
satisfied by ${\bf s}_t({\bf x})$, ${\bf s}_l({\bf x})$, and $T(t)$ are
(\ref{A5}) and (\ref{A6}):

\begin{mathletters}
 \label{helmt}
  \begin{eqnarray}
   &\nabla^2{\bf s}_t+{\cal K}^2\,{\bf s}_t=0 & \label{helmt.a} \\
   &\mu\,T+\mu'\,\dot T+{\cal K}^{-2}\ddot T=0 & \label{helmt.b}
  \end{eqnarray}
\end{mathletters}
and

\begin{mathletters}
 \label{helml}
  \begin{eqnarray}
   &\nabla^2{\bf s}_l+{\cal Q}^2\,{\bf s}_l=0 & \label{helml.a} \\
   &\mu\,T+\mu'\,\dot T+{\cal Q}^{-2}\ddot T=0 & \label{helml.b}
  \end{eqnarray}
\end{mathletters}

Since $T(t)$ must fulfil {\it both\/} equations (\ref{helmt.b}) and
(\ref{helml.b}), the {\it separation constants} ${\cal K}^2$ and ${\cal Q}^2$
are {\it not\/} independent. The binding relationship is established after
it is realised that the solution to those equations is of the form

\begin{equation}
      T(t) = e^{\gamma t}     \label{Tt}
\end{equation}
where $\gamma\/$ is of course a {\it complex\/} quantity. We readily obtain

\begin{mathletters}
 \label{KQ}
  \begin{eqnarray}
   {\cal Q}^2 & \!= \! & -\rho\gamma^2\left[
   \lambda+2\mu+\gamma\,(\lambda'+2\mu')\right]^{-1}  \label{KQ.a} \\
   {\cal K}^2 & \!= \! & -\rho\gamma^2\left
   [\mu+\gamma\,\mu'\right]^{-1} \label{KQ.b}
  \end{eqnarray}
\end{mathletters}

The values $\gamma\/$ can possibly take on are determined by the {\it boundary
conditions\/}, equations (\ref{I.4}). The reader is referred to Appendix B
for a detailed description of the eigenvalue algebra of this problem. Just as
in the non-dissipative case, there are seen to be two families of eigenmodes:
{\it toroidal\/} (purely torsional) and {\it spheroidal\/}. Viscous effetcs
are however small in practice, as inferred from the narrow {\it linewidth\/}
of the measured resonances. This means that the following inequalities hold
in any cases of interest to us:

\begin{equation}
  \frac{\mu'}{\mu},\;\frac{\lambda'}{\mu}\ll\frac{1}{\omega}
  \label{approx}
\end{equation}
where $\omega\/$ is the frequency of the mode considered. We shall use these
inequalities to estimate the roots of the eigenvalue equation (\ref{eigen})
(see Appendix B) {\it perturbatively\/} from the non-dissipative ones,
already known ---see \cite{lobo} for full details. Clearly thus, our
procedure will be valid for the lower frequency modes. We proceed
sequentially for the two families of eigenmodes.

\subsection{Toroidal modes}

These correspond to the solutions to (see (\ref{eigen}))

\begin{equation}
  \beta_1({\cal K}R) = 0   \label{torKV}
\end{equation}

which is {\it formally\/} identical to the toroidal eigenvalue equation
for a non-dissipative solid sphere \cite{lobo}. If we call $k_{nl}^T$
the toroidal wave numbers of the latter, we have

\begin{equation}
  {\cal K}_{nl}^T = k_{nl}^T = \sqrt{\frac{\rho}{\mu}}\omega_{nl}^T
  \label{KRTKV}
\end{equation}
and hence, by equation (\ref{KQ.b}),

\begin{equation}
  \gamma_{nl}^T = -(\omega_{nl}^T)^2\frac{\mu'}{2\mu} +
  i\omega_{nl}^T\sqrt{1-\left(\frac{\omega_{nl}^T\mu'}{2\mu}\right)}
  \simeq i\omega-\omega^2\frac{\mu'}{2\mu}
  \label{gammaT}
\end{equation}
where the last approximation depends on the validity of the assumption
(\ref{approx}). Expression (\ref{gammaT}) nicely shows how this Kelvin-Voigt
model predicts {\it exponentially damped\/} eigenmode oscillations. If we
recall that such damping is expediently described in terms of a {\it quality
factor Q\/} ---see (\ref{I.1}) above--- then we discover that the prediction
of the model is that

\begin{equation}
  Q_{nl}^T = \frac{2\mu}{\mu'}\left(\omega_{nl}^T\right)^{-1}  \label{Qtor}
\end{equation}
i.e., the quality factor for toroidal modes is {\it inversely proportional\/}
to the frequency of the mode. The {\it amplitudes\/} of these modes have the
form

\begin{equation}
  {\bf s}^{T}_{KV}({\bf x},t)={\bf s}^{T}_{E}({\bf x},t)\,e^{-\omega t/Q}\ ,
  \qquad Q=\frac{2\mu}{\mu'\omega}
\end{equation}
where the subindex $KV\/$ stands for `Kelvin-Voigt', while $E\/$ refers
to the standard frictionless case, whose amplitudes are those given e.g.
in reference \cite{lobo}.

\subsection{Spheroidal modes}\label{qnsphm}

A second alternative to find a non-trivial solution of the linear system
(\ref{bcatlastKV}) is to impose the condition

\begin{equation}
 \beta_4\left({\cal Q} R,\frac{\lambda+\gamma\lambda'}{\mu+\gamma\mu'}\right)
 \,\beta_3({\cal K} R)-l(l+1)\,\beta_1({\cal Q} R)\beta_1({\cal K} R) = 0
 \label{sphKV}
\end{equation}

This is characteristic of the {\it spheroidal\/} eigenmodes. By virtue of
equations (\ref{KQ}), this relationship can be translated into a condition
to be fufilled by $\gamma$, and which depends on the ratios $\lambda/\mu$,
$\lambda'/\lambda$ and $\mu'/\mu$, as well as on the multipole index $l\/$.
In this case, as we are not dealing with an eigenvalue problem of a
selfadjoint operator, complex solutions to equation (\ref{sphKV}) are
allowed ---indeed, expected. An exact solution of that equation implies
a separation of its real and imaginary parts, followed by numerical
calculations which determine the angular frequency and quality factor
of the quasinormal mode at hand. We are interested in materials with long
decay times, so we shall set up a perturbative solution to equation
(\ref{sphKV}), using 

\begin{equation}
  \epsilon\equiv\frac{\mu'\omega}{\mu}   \label{eps}
\end{equation}
as the small perturbative parameter. In other words, we assume that the
approximation (\ref{approx}) holds. Here, $\omega\/$ stands for a generic
spheroidal eigenfrequency of the non-dissipative solid. Obviously, the
unperturbed solution, i.e., that corresponding to $\epsilon\/$\,=\,0, is
the elastic solid's solution, already discussed in reference \cite{lobo}.
We thus introduce the perturbative expansion

\begin{equation}
  \gamma=\gamma_0+\gamma_1\epsilon+O(\epsilon^2),\hspace{1cm}
  \gamma_0=-i\omega,\label{gamma}
\end{equation}

Using equations (\ref{KQ}) and (\ref{gamma}), we obtain perturbative
expansions for the parameters ${\cal K}$ and ${\cal Q}$ which can be
written as

\begin{equation}
  {\cal K}=k_0+k_1\epsilon+O(\epsilon^2),\qquad
  {\cal Q}=q_0+q_1\epsilon+O(\epsilon^2),\label{perkq}
\end{equation}
where

\begin{mathletters}
\label{kq}
\begin{eqnarray}
  & k_0=k=\omega\sqrt{\frac{\rho}{\mu}}\ ,\qquad
    k_1=i\sqrt{\frac{\rho}{\mu}}\left(\gamma_1+\frac{\omega}{2}\right) &
    \label{kq.a} \\[1 em]
  & q_0=q=\omega\sqrt{\frac{\rho}{\lambda+2\mu}}\ ,\qquad
    q_1=i\sqrt{\frac{\rho}{\lambda+2\mu}}\left(\gamma_1+
    \frac{h'+2}{h+2}\frac{\omega}{2}\right) &  \label{kq.b}
\end{eqnarray}
\end{mathletters}

In the above equations, $k$ and $q$ are the parameters appearing in the
elastic sphere's case, and we have introduced the dimensionless ratios

\begin{equation}
  h\equiv\frac{\lambda}{\mu}\ ,\qquad h'\equiv\frac{\lambda'}{\mu'}
  \label{hs}
\end{equation}
which are both zero order quantities. We can now perform the perturbative
expansion of the eigenvalue equation. In order to ease the resulting
expressions, let us introduce the following notation:

\begin{eqnarray}
 l(l+1)\,\beta_1({\cal K} R) & = & l(l+1)\beta_1(kR)+l(l+1)
 \beta'_1(kR)k_1R\epsilon  \nonumber  \\
 & \equiv & B_0+B_1k_1R\,\epsilon  \\
 \beta_1({\cal Q} R) & = & \beta_1(qR)+\beta'_1(qR)q_1R\,\epsilon
 \equiv C_0+C_1q_1R\,\epsilon  \\
 \beta_3({\cal K} R) & = & \beta_3(kR)+\beta'_3(kR)k_1R\,\epsilon
 \equiv D_0+D_1k_1R\,\epsilon  \\
 \beta_4\left({\cal Q} R,\frac{\lambda+\gamma\lambda'}{\mu+\gamma\mu'}\right)
 & = & \beta_4(qR,h)+\left[\beta'_4(qR,\lambda/\mu)q_1R-\frac{i}{2}
 (h'-h)\,j_l(qR)\right]\epsilon  \nonumber  \\
 & \equiv & A_0+(A_1q_1R-iA'_1)\,\epsilon
\label{bet4}
\end{eqnarray}
where a prime over a $\beta\/$ function denotes differentiation with respect
to its {\it first\/} argument. We note that the uppercase constants
introduced above are real. With this notation, the zeroth order form of
equation (\ref{sphKV}) is

\begin{equation}
   A_0D_0-C_0B_0=0
\end{equation}
which is simply the condition that $\omega\/$ be a spheroidal eigenvalue of
the purely elastic case. On the other hand, the first order expansion of
(\ref{sphKV}) yields

\begin{equation}
   (A_0D_1-C_0B_1)k_1R+(A_1D_0-C_1B_0)q_1R=iA'_1D_0,
\end{equation}
whence, using the form of $k_1$ and $q_1$, the value of $\gamma_1$ ensues.
It can be written as

\begin{equation}
\gamma_1=-\frac{\omega}{2}f(kR,h,h'),\label{vgam}
\end{equation}
where the dimensionless function $f\/$ has the form\footnote{
The case in which $f\/$ takes its simplest form is that of monopole modes.
We know (see \cite{lobo}) that when $l\/$\,=\,0 equation (\ref{sphKV}) is
no longer valid, and must be replaced by

\begin{equation}
  \beta_4\left({\cal Q} R,\frac{\lambda+\gamma\lambda'}{\mu+\gamma\mu'}
  \right) = 0
\end{equation}

Using now the expansion (\ref{bet4}) and the fact that $A_0=0$ for the
uperturbed monopole eigenfrequencies, we obtain for the first order
correction to ${\cal Q}$,

\begin{equation}
  q_1=iA'_1A_1^{-1}\label{q1l0}
\end{equation}
and therefore, using the notation of equation (\ref{vgam}) and the relation
(\ref{kq}):

\begin{equation}
f(qR,h,h')=\frac{h'+2}{h+2}-\frac{2\,A'_1}{qR\,A_1}.
\end{equation}}

\begin{equation}
  f(kR,h,h') = -\frac{2A'_1D_0(kR)^{-1}-A_0D_1+C_0B_1-(A_1D_0-C_1B_0)
  \,(h'+2)(h+2)^{-3/2}}{A_0D_1-C_0B_1+(A_1D_0-C_1B_0) (h+2)^{-1/2}}
  \label{f}
\end{equation}

We note that the first order correction obtained for $\gamma$ is {\it real}.
Therefore, to this order of approximation, the frequencies of vibration
remain unaltered, and are the same as those obtained for the elastic solid.
Moreover, $k_1$ and $q_1$ happen to be purely imaginary. Therefore, the
modulus of the radial functions appearing in the spatial part of spheroidal
quasinormal modes of vibration will also be the same as those of the
elastic solid, for the corrections to $k\/$ and $q\/$ will just introduce,
to first order, a {\it complex phase\/} factor.

\begin{figure}[ht]
\psfig{file=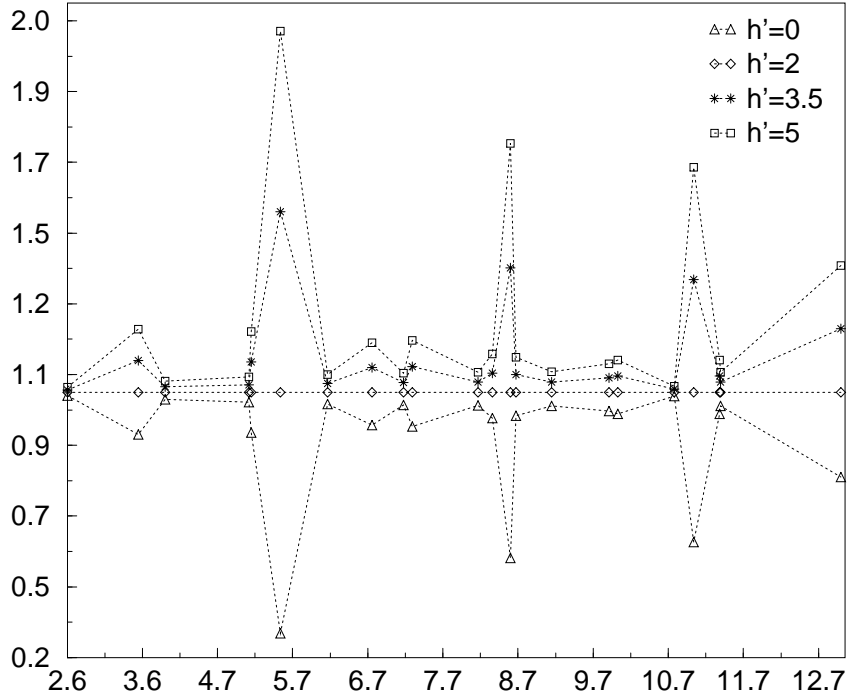,height=12cm,width=12cm,rheight=10cm,bbllx=-3cm,bblly=-1cm,bburx=14.2cm,bbury=16.0cm}
\caption{Plot of the function $f(kR,h,h')$ ---see (\protect\ref{f})---
for $h\/$\,=\,2 and a Poisson ratio $\sigma\/$\,=\,1/3, for the first 20
(spheroidal) modes of the sphere's spectrum and a few values of the ratio
$h'\/$.   \label{f1}}
\end{figure}

\begin{figure}[htb]
\psfig{file=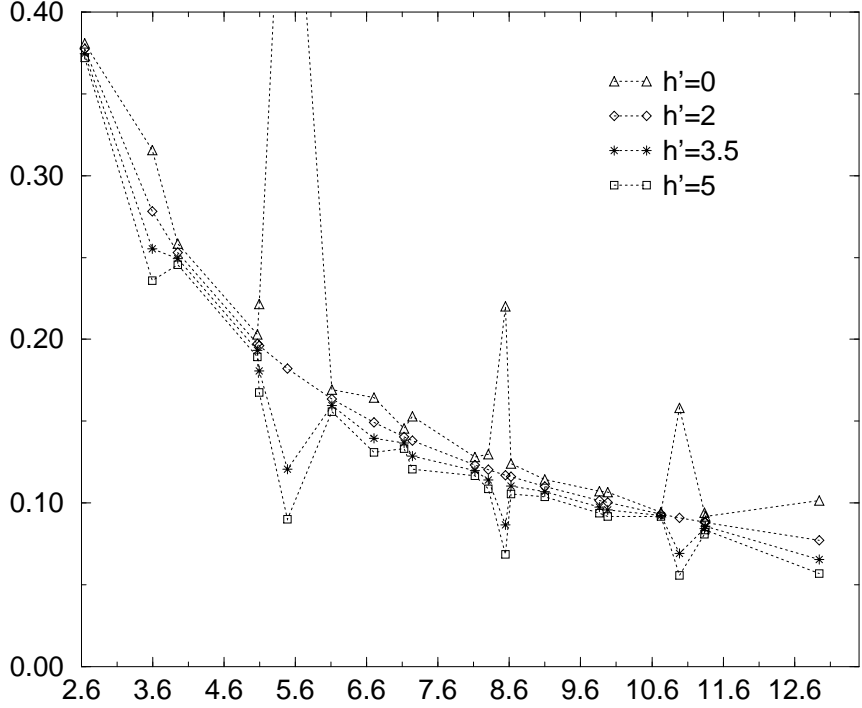,height=12cm,width=12cm,rheight=10cm,bbllx=-3cm,bblly=-1cm,bburx=14.2cm,bbury=16.0cm}
\caption{Quality factor $Q_{nl}\/$ (in units of $[\mu R/\mu'v_t]$) for a few
values of the parameter $h'\/$ and the first modes of the spheroidal spectrum.
The viscoelastic solid is described by a Kelvin-Voigt model with $h\/$\,=\,2.
\label{f2}}
\end{figure}

Summing up, we have shown that while the spheroidal normal modes of vibration
of an elastic solid are given by an expression of the form \cite{lobo}

\begin{equation}
  {\bf s}_E^P({\bf x},t)=e^{i\omega_{nl}^Pt}\left[
  A_nl(r)\,Y_{lm}(\theta,\varphi)\,{\bf n}
  -B_{nl}(r)\,i{\bf n}\!\times\!{\bf L}Y_{lm}(\theta,\varphi)\right],
  \label{sEP}
\end{equation}
the spheroidal quasinormal modes ${\bf s}_{KV}^P$ of a Kelvin--Voigt solid
are obtained from the normal modes of the elastic solid according to the
following

\begin{equation}
  {\bf s}_{KV}^P({\bf x},t) = e^{i\omega_{nl}^Pt-\omega_{nl}^Pt/Q}
  \,\left[e^{i\chi_1(r)}A_{nl}(r)Y_{lm}(\theta,\varphi)\,{\bf n} -
  e^{i\chi_2(r)}B_{nl}(r)\,i{\bf n}\!\times\!{\bf L}Y_{lm}(\theta,\varphi)
  \right]   \label{qnKV}
\end{equation}
the qualitity factor being given by

\begin{equation}
  Q_{nl} = \frac{2\mu}{\mu'\omega_{nl}^P}\,\frac{1}{f}  \label{QQnKV}
\end{equation}
where, it is recalled, $f\/$ is given by (\ref{f}) as a function of the mode
and the coefficients characterising the viscoelastic properties of the solid.
The real phases $\chi^{}_{1,2}(r)$ can be computed from equations
(\ref{sKV.1}), (\ref{sKV.2}) and (\ref{kq}). Nevertheless the explicit (and
cumbersome) form of these phases is largely irrelevant and whe shall omit
its explicit form here \cite{jup}. They merely introduce a position dependent
shift in the phase of the vibrations which is of order $\epsilon\/$, therefore
not likely to give rise to measurable effects. More interesting and physically
relevant is the behaviour of the function $f\/$ giving the precise dependence
of the quality factor on frequency. First of all, it is easily seen that, for
the special case $h\/$\,=\,$h'\/$, $f\/$ is equal to 1, and thus the quality
factor is proportional to $\omega^{-1}$, as was the case with toroidal modes.
But when the aforementioned equality does not hold, numerical calculations
are needed. Figure \ref{f1} shows that this function is not very strongly
dependent on frequency, though appreciable deviations from constant behaviour
can be seen at different modes. We have represented $f\/$ for the first 20
eigenvalues of the spheroidal spectrum, where the inequality (\ref{approx})
safely holds. It is clear from equation (\ref{f}) that holding fixed $h\/$
and $kR\/$ leaves us with a {\it linear\/} function of $h'\/$, whose slope
varies from root to root. Figure \ref{f2} displays the {\it quality factor}
$Q\/$ for the same set of eigenvalues. To be observed is the global trend
towards an $\omega^{-1}\/$ dependence of $Q\/$ on $\omega\/$, which exactly
holds for the toroidal modes ---cf.\ (\ref{Qtor}) above. Fluctuations are
however present in certain modes for particular values of the ratio $h'\/$
of bulk to shear viscosity coefficients as a consequence of the variability
of the $f\/$ function.

\section{Maxwell model}

In this section we shall consider constituent equations given by the so
called Maxwell model. Like the Kelvin-Voigt, these equations only involve
first order time derivatives. As we shall see, quite different predictions
will be obtained for the quality factor dependence on the mode frequencies.

The Maxwell model is also isotropic and homogenous, and is characterised by
the following constituent equations \cite{ja}:

\begin{equation}
  \partial_t\sigma_{ij}
  +\alpha\,\sigma_{kk}\delta_{ij}^{}+\beta\,\sigma_{ij} =
  \frac{\partial}{\partial t}\,\left(\lambda\,s_{kk}\,\delta^{}_{ij}
  + 2\mu\,s_{ij}\right)    \label{constM}
\end{equation}

Here, the constants $\lambda$ and $\mu$ are again the Lam\'e coefficients
describing the elastic behaviour of the body, while the constants $\alpha$
and $\beta\/$ parametrise the effects due to internal friction. To construct
factorised solutions, we must also factorise both stress and strain\footnote{
Due to the equations of motion (\ref{eqmov}), if ${\bf s}({\bf x},t)$ is
assumed to be separable in the fashion (\ref{separ}) then the strain
tensor is separable too, and due to the constituent equation it is easily
seen that the only possible time dependence is of the form 
\mbox{$\exp(\gamma t)$}.}:

\begin{equation}
  \sigma_{ij}({\bf x},t)=e^{\gamma t}\,\sigma_{ij}({\bf x})\ ,  \qquad
  s_i({\bf x},t) = e^{\gamma t}\, s_i({\bf x}),
\end{equation}

The constituent equation is thus written, after separation of variables
and contraction of its free indices, as

\begin{equation}
  (\gamma+3\alpha+\beta)\,\sigma_{jj}({\bf x}) =
  \gamma\,(2\mu+3\lambda)\,s_{jj}({\bf x})
\end{equation}

and hence we have the following relationship between the spatial parts of
stress and strain:

\begin{equation}
  \left(1+\frac{\beta}{\gamma}\right)\,\sigma_{ij}({\bf x}) =
  \left(\lambda-\alpha\frac{2\mu+3\lambda}{\gamma+3\alpha+\beta}\right)
  \,s_{kk}({\bf x}) \delta_{ij}^{}+2\mu\,s_{ij}({\bf x})
\end{equation}

Like in the Kelvin-Voigt model, we shall be mainly interested in the case
of small internal friction, so we shall assume

\begin{equation}
  \frac{\beta}{\mid\gamma\mid},\;\frac{\alpha}{\mid\gamma\mid}\ll 1
  \label{ApM}
\end{equation}
whence the following constituent relationship results:

\begin{equation}
  \sigma_{ij}({\bf x})=\lambda\left(1-\frac{\delta}{\gamma}\right)
  \,s_{kk}({\bf x})\delta^{}_{ij} + 2\mu\left(
  1-\frac{\beta}{\gamma}\right)\,s_{ij}({\bf x})     \label{consM}
\end{equation}
where we have introduced a new constant, $\delta$, defined by

\begin{equation}
  \delta\equiv\frac{2\mu+3\lambda}{\lambda}\,\alpha + \beta,
\end{equation}
so that we can take as the parameters characterizing the Maxwell solid
the set consisting of the Lam\'e coefficients $\lambda$, $\mu$, and the
parameters $\beta$, $\delta\/$ which describe internal friction.

Let us compare equation (\ref{consM}) with that of the Kelvin--Voigt model
---cf. (\ref{consKV})--- once the separation of variables has been performed:

\begin{equation}
  \sigma_{ij}({\bf x})=(\lambda+\gamma\lambda')\,s_{kk}({\bf x})\,\delta_{ij}
  + 2(\mu+\gamma\mu')\,s_{ij}({\bf x})    \label{consKVag}
\end{equation}

Comparing equations (\ref{consM}) and (\ref{consKVag}), we observe that the
solution of the Maxwell model can be carried out, as regards the spatial
part of ${\bf s}$, following the same method used in the previous section
for the Kelvin--Voigt model. In fact, we can directly take the expressions
there derived, and make the substitutions

\begin{equation}
  \mu'\ \longrightarrow\ -\mu\beta\gamma^{-2}\ ,\qquad
  \lambda'\ \longrightarrow\ -\lambda\delta\gamma^{-2}
\end{equation}
which transform equation (\ref{consKVag}) into (\ref{consM}). Thus the form
of the solutions and boundary conditions for a Maxwell viscoelastic sphere
are those of Appendix \ref{aB} with constants ${\cal K}\/$ and ${\cal Q}\/$
now given by the following functions of the parameter $\gamma$:

\begin{mathletters}
 \label{KQM}
 \begin{eqnarray}
  {\cal Q} & = & i\sqrt{\frac{\rho}{\lambda+2\mu}}\left[\gamma+\frac{1}{h+2}
  \left(\frac{h}{2}\delta+\beta\right)\right]  \label{KQM.a}  \\
  {\cal K}&=&i\sqrt{\frac{\rho}{\mu}}\left(\gamma+\frac{\beta}{2}\right)
  \label{KQM.b}
 \end{eqnarray}
\end{mathletters}
where the approximation (\ref{ApM}) has been taken into account, and
$h\/$\,$\equiv$\,$\lambda/\mu$.

Thus the two families of quasinormal modes of vibration are also present
in this model, and we describe them in the following subsections.

\subsection{Toroidal modes}

As already discussed, the allowed values for $\gamma$ are again those
making the linear system (\ref{bcatlastKV}) compatible, and there are two
alternative ways to accomplish this. The first possibility yields purely
tangential ($C_t=C_l=0$) vibrations satisfying once more the condition

\begin{equation}
  \beta_1({\cal K} R) = 0\ \ \Longrightarrow\ \ 
  {\cal K} = \sqrt{\frac{\rho}{\mu}}\,\omega_{nl}^T
\end{equation}
$\omega_{nl}^T$ being a toroidal eigenfrequency of the elastic sphere. Using
the relationship between $\gamma$ and ${\cal K}$ for a Maxwell solid given
by equation (\ref{KQM}), we obtain the allowed values $\gamma_{nl}^T$ as

\begin{equation}
  \gamma_{nl}^T = -i\omega_{nl}^T-\frac{\beta}{2}    \label{gammaM}
\end{equation}

Again, toroidal quasinormal modes have two fundamental properties: they
have the same set of eigenfrequencies as the elastic sphere (to first order
in the parameters describing internal friction, $\beta$ in this case), and
also exactly the same spatial part (for {\em all\/} values of the viscosity
parameters). The only difference between Kelvin--Voigt and Maxwell solids as
regards toroidal modes appears in the dependence of the quality factor on
$\omega$: as equation (\ref{gammaM}) shows, the quality factor in a linear
Maxwell solid increases linearly with frequency. We can express all these
properties by means of the following formul\ae:

\begin{equation}
  {\bf s}_{M}^T({\bf x},t) = {\bf s}_{E}^T({\bf x},t)
  \,e^{-\omega_{nl}^Tt/Q}\ ,\qquad  Q_{nl} = \frac{2\omega_{nl}^T}{\beta}
  \label{4.13}
\end{equation}
relating Maxwell quasinormal modes of vibration, ${\bf s}_M^T({\bf x},t)$,
to elastic normal modes, ${\bf s}_E^T({\bf x},t)$, for the toroidal family.

\begin{figure}[t]
\psfig{file=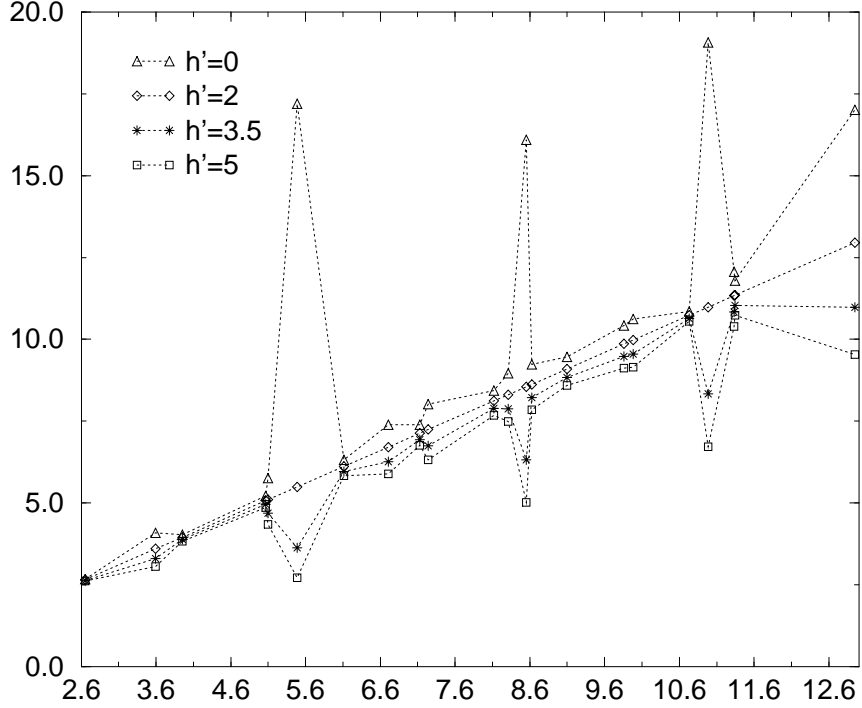,height=12cm,width=12cm,rheight=10cm,bbllx=-3cm,bblly=-1cm,bburx=14.2cm,bbury=16.0cm}
\caption{Quality factor $Q_{nl}\/$ (in units of $[v_t/\beta R]$) for a few
values of the parameter $h'\/$ and the first modes of the spheroidal spectrum.
The viscoelastic solid is described by a Maxwell model with $h\/$\,=\,2.
\label{f3}}
\end{figure}

\subsection{Spheroidal modes}

In order to handle the spheroidal family, we shall resort again to the
perturbative expansions already used in the Kelvin--Voigt case, and also
in the toroidal family, just described. The Maxwell model trivially reduces
to the perfect elastic case when $\beta\/$\,=\,$\delta$\,=\,0, hence we can
take as the perturbative parameter

\begin{equation}
  \epsilon = \frac{\beta}{\omega}
\end{equation}
where $\omega\/$ is the elastic eigenfrequency to which $\gamma\/$ approaches
when both $\beta\/$ and $\delta\/$ approach zero. Perturbative expansions in
the fashion of section \ref{qnsphm} can now be introduced:

\begin{equation}
  \gamma = -i\omega+\gamma_1\epsilon\ ,\qquad
  {\cal K} = k+k_1\epsilon\ ,\qquad
  {\cal Q}=q+q_1\epsilon
\end{equation}
where the first order corrections $k_1$ and $q_1$ are given by equations
(\ref{KQM}) as functions of $\gamma_1$:

\begin{mathletters}
 \label{kqM}
 \begin{eqnarray}
  k_1 & = & i\sqrt{\frac{\rho}{\mu}}\left(\gamma_1+\frac{\omega}{2}\right)
   \label{kqM.a}  \\
  q_1 & = & i\sqrt{\frac{\rho}{\lambda+2\mu}}\left(\gamma_1 +
          \frac{h'+2}{h+2}\frac{\omega}{2}\right)
   \label{kqM.b}
 \end{eqnarray}
\end{mathletters}

The zero--order ratio $h'\/$ is now given by

\begin{equation}
  h' = \frac{\alpha}{\beta}\,h
\end{equation}

With this definition, together with that of the perturbative parameter,
the expressions at hand are formally identical to those of the Kelvin--Voigt
model, and therefore the solutions to the Maxwell model share all their
properties with their Kelvin--Voigt counterparts; the exception is the
dependence of the quality factor on frequency: the product $\gamma_1\epsilon$,
which gives the exponential decay, is now {\it independent of\/} $\omega$. 

Summing up, spheroidal quasinormal modes of the Maxwell solid,
${\bf s}_{M}^P({\bf x},t)$, are related to spheroidal normal modes of a
perfectly elastic sphere by the equations

\begin{equation}
  {\bf s}_{M}^P({\bf x},t) = {\bf s}_{E}^P({\bf x},t)
  \,e^{-\omega_{nl}^Pt/Q+\chi(r)}\ ,\qquad
  Q_{nl} = \frac{2\omega_{nl}^P}{\beta}\,\frac{1}{f}
  \label{qnmM}
\end{equation}
where the function $f(kR,h,h')$ is again given by (\ref{f}). In Figure
\ref{f3} we plot the quality factor of the first twenty eigenmodes of a
Maxwell solid. We note that $Q_{nl}\/$ is proportional to the eigenmode
frequency $\omega^P_{nl}\/$, just like in the toroidal modes ---cf.
(\ref{4.13})---, but fluctuations around this behaviour are oberved for
certain modes which are associated to corresponding ones in $f\/$.

As we see, the only difference between the behaviour of Maxwell and
Kelvin--Voigt viscoelastic solids, when the internal friction effects can
be considered small, appears in the dependence of $Q\/$ on frequency. We
must however stress that, under other conditions (e.g. static load), both
models show larger divergences in their physical properties~\cite{GI}.

\section{Other Models}

In this section we review other models which have been proposed to address
the dynamics of a viscoelastic solid. They are generalisations of those in
the two previous sections. We shall however not attempt to find complete
solutions to all of them, as it eventually becomes too cumbersome. We shall
however discuss in this section some of their most relevant traits.

\subsection{The Standard Linear Model}

The Standard Linear Model (SLM) for a viscoelastic solid is a generalised
combination of the Kelvin--Voigt and Maxwell models. The constituent
equations take here the form:

\begin{equation}
  \sigma_{ij} + \frac{\partial}{\partial t}\,(\alpha\,\sigma_{kk}\delta_{ij}
  + 2\beta\,\sigma_{ij}) = \left(\lambda+\alpha'\frac{\partial}{\partial t}
  \right)\,s_{kk}\delta_{ij} + 2\,\left(\mu+\beta'\frac{\partial}{\partial t}
  \right)\,s_{ij}    \label{consSLM}
\end{equation}
where the effects of internal friction are described in this case with the
aid of four constant parameters: $\alpha$, $\beta$, $\alpha'$, and $\beta'$.
When looking for factorised solutions, the equations of motion and the above
relationship force a time dependence of the form $e^{\gamma t}$ for both
stress and strain. When such a dependence is introduced in equation
(\ref{consSLM}), we obtain the following relationship between the spatial
parts of the stress and strain tensors:

\begin{equation}
  (1+2\gamma\beta)\sigma_{ij}({\bf x}) = \left[\lambda+\alpha'\gamma -
  \alpha\gamma\frac{3\lambda+2\mu+\gamma(3\alpha'+2\beta)'}{1+\gamma
  (3\alpha+2\beta)}\right]s_{kk}({\bf x})\delta_{ij}+2(\mu+\beta'\gamma)
  \,s_{ij}({\bf x})
\end{equation}

The case of small internal friction is treated by first order approximation
in the quantities parametrising viscous processes, i.e.,

\begin{equation}
  \alpha,\;\beta,\;\alpha',\;\beta'\;\ll\;\mid\gamma\mid^{-1}
  \label{little}
\end{equation}

When such approximation is made, the above equation reduces to

\begin{equation}
  \sigma_{ij}({\bf x}) = (\lambda+\lambda'\gamma)s_{kk}({\bf x})\delta_{ij}
  + 2(\mu+\mu'\gamma)s_{ij}({\bf x})
\end{equation}
where we have introduced two new constants given by

\begin{equation}
  \mu'\equiv\beta'-2\beta\mu\ ,\qquad
  \lambda'\equiv\alpha'-2\lambda\beta - (3\lambda+2\mu)\alpha
  \label{lm}
\end{equation}

Therefore when equation (\ref{little}) holds, the SLM reduces to a
Kelvin--Voigt model, i.e., for small internal friction, both models have
the same set of quasinormal modes of vibration, which are characterized
by the constants $\lambda$, $\mu$, $\lambda'$ and $\mu'$, the latter being
given, for the Standard Linear solid, by equations (\ref{lm}).

\subsection{Generalised mechanical models}

The models analyzed so far are the simplest ones obtained by three dimensional
generalisations of mechanical viscoelastic models composed of linear springs.
They give rise to differential constituent relations, with time derivatives
up to the first order. Considering more involved generalisations yields
differential relations involving higher order time derivatives of strain and
stress ---see e.g. \cite{HA}. Thus, quite independently of any reference to
the underlying mechanical model, we can consider general differential
relations between stress and strain including any number of time derivatives.
To ease the formulation of such differential constituent equations for the
case of isotropic and homogenous solids, we shall introduce the trace-free
parts of the strain and stress tensors, $s'_{ij}$, $\sigma'_{ij}$ (usually
termed {\em deviatoric\/} components in the literature on viscoelasticy
\cite{HA}), and their traces, $s\/$ and $\sigma\/$ ({\em dilational\/}
components), defined by

\begin{mathletters}
\begin{eqnarray}
  s'_{ij} & = & s_{ij} - \frac{1}{3}\,s\,\delta_{ij}\ ,\qquad s = s_{kk} \\
  \sigma'_{ij} & = & s_{ij} - \frac{1}{3}\sigma\delta_{ij}\ ,\qquad
  \sigma = \sigma_{kk}.
\end{eqnarray}
\end{mathletters}

In terms of the above quantities, the linear Hooke law for an elastic solid
takes the form

\begin{equation}
  \sigma = (3\lambda+2\mu)s\hspace{1cm}\sigma'_{ij} =
  2\mu\,s'_{ij}   \label{conselast}
\end{equation}
while the constituent equation of an SLM is written

\begin{equation}
  \left[1+(3\alpha+2\beta)\,\frac{\partial}{\partial t}\right]\,\sigma =
  \left[(3\lambda+2\mu) + (3\alpha'+2\beta')\,
  \frac{\partial}{\partial t}\right]\,s\ ,\qquad
  \left(1+2\beta\,\frac{\partial}{\partial t}\right)\,\sigma'_{ij} =
  \left(2\mu+2\beta'\,\frac{\partial}{\partial t}\right)\,s'_{ij}
\end{equation}

This equation can now be generalised to include higher order time
derivatives. We can thus consider viscoelastic models whose constituent
equation is given by

\begin{equation}
  R(\partial_t)\sigma = S(\partial_t)s\ ,\qquad
  R'(\partial_t)\sigma'_{ij} = S'(\partial_t)s'_{ij}   \label{consgen}
\end{equation}
where we have introduced the polynomials

\begin{equation}
   R(x) \equiv\sum_{l=0}^{N-1}r_l x^l\ ,\qquad
   R'(x)\equiv\sum_{l=0}^{N-1}r'_lx^l\ ,\qquad
   S(x) \equiv\sum_{l=0}^{N-1}s_l x^l\ ,\qquad
   S'(x)\equiv\sum_{l=0}^{N-1}s'_lx^l   \label{poly}
\end{equation}
so that a general differential model is given for each set of 4$N\/$ real
constants $r_l\/$, $r'_l\/$, $s_l\/$, and $s'_l\/$ characterising the solid.
Some of these constants may vanish. The Kelvin-Voigt, Maxwell and SL models
considered above are of course special cases within this general class.
Several procedures have been proposed in the literature to solve the general
equations ---see \cite{MA} for a review and further reference---, which can
be applied to the solid viscoelastic sphere problem. We now sketch how they
work in this case of our interest.

Let $\tilde\sigma_{ij}({\bf x},\Omega)$ and $\tilde s_{ij}({\bf x},\Omega)$
be the Fourier transforms of the stress and strain tensors, and
$\tilde s_i({\bf x},\Omega)$ that of the displacement vector field:

\begin{mathletters}
 \label{ft}
 \begin{eqnarray}
  \tilde\sigma_{ij}({\bf x},\Omega) & \equiv &
     \int_{-\infty}^\infty\sigma_{ij}({\bf x},t)e^{i\Omega t}\,dt
     \label{ft.a}  \\
  \tilde s_i({\bf x},\Omega) & \equiv &
     \int_{-\infty}^\infty s_{i}({\bf x},t)e^{i\Omega t}\,dt
     \label{ft.b}  \\
  \tilde s_{ij}({\bf x},\Omega) & \equiv &
     \int_{-\infty}^{\infty}s_{ij}({\bf x},t)e^{i\Omega t}\,dt
     \label{ft.c}
 \end{eqnarray}
\end{mathletters}

In terms of these, Fourier transforms the constituent equations
(\ref{consgen}) read:

\begin{equation}
  \tilde\sigma = \frac{R(i\Omega)}{S(i\Omega)}\,\tilde s\ ,\qquad
  \tilde\sigma'_{ij} = \frac{R'(i\Omega)}{S'(i\Omega)}\,\tilde s_{ij}
  \label{constran}
\end{equation}
while the equations of motion are 

\begin{equation}
  -\Omega^2\tilde{\bf s} = \frac{1}{3}\left(\frac{R(i\Omega)}{S(i\Omega)} -
   \frac{R'(i\Omega)}{S'(i\Omega)}\right)\nabla(\nabla{\bf \cdot}
   \tilde{\bf s}) + \frac{R'(i\Omega)}{2S'(i\Omega)}\,\nabla^2\tilde{\bf s}
\end{equation}

Comparing the above equations with the corresponding ones for normal modes
of vibration of elastic solids, and the constituent relation (\ref{constran})
with (\ref{conselast}), we note that the problem of finding solutions to the
equation of motion of a general viscoelastic differential model reduces to
that of finding the normal modes of vibration of an elastic solid having
{\it complex\/} Lam\'e coefficients given by

\begin{equation}
  \tilde\lambda(\Omega) = \frac{1}{3}\,\left(\frac{R(i\Omega)}{S(i\Omega)}
  - \frac{R'(i\Omega)}{S'(i\Omega)}\right)\ ,\qquad
  \tilde\mu(\Omega) = \frac{1}{2}\frac{R'(i\Omega)}{S'(i\Omega)}
\end{equation}
where the allowed values of $\Omega$ are obtained as the solutions to the
elastic solid's eigenfrequency equation when the above complex coefficients
are used instead of the real, constant Lam\'e coefficients $\lambda$, $\mu$.
Generally, $\Omega$ will have complex values, thus giving rise to damped
system oscillations. After solving for $\Omega$, the spatial part of the
solutions is obtained from that of the normal modes by simply substituting
the old, real--valued constants $\omega$, $\lambda$ and $\mu$ by the new
complex values $\Omega$, $\tilde\lambda$ and $\tilde\mu$. This method for
solving the viscoelasticity is often termed in the literature on the subject
the {\em Correspondence Principle} \cite{MA}, and as a matter of fact our
previous derivations of the form of the quasinormal modes for Kelvin-Voigt,
Maxwell and SL models can be seen to be special cases of its application.
The method is applicable to any boundary value problem whose elastic
counterpart is solvable. The case of small internal friction (i.e., first
order approximation in the coefficients of the polynomials (\ref{poly}) has
been considered by Graffi \cite{MA} for one dimensional wave propagation.

The three dimensional spherical case is also solvable, as we know. The
{\it toroidal\/} modes are relatively straightforward to obtain from their
elastic counterparts due to the simple form of their eigenvalue equation,
while the spheroidal ones demand more complex algebra, which becomes
increasingly cumbersome as the order $N\/$ of the model increases.
We shall present here the general solution for the toroidal modes for any
differential viscoelastic model, whereby we shall obtain the dependence of
their $Q\/$ on frequency. This will also be the approximate dependence for
the spheroidal modes, if friction effects are small, as was the case with
the first order models analysed so far. A complete solution for the latter
modes can also be systematically found, but will be omitted due to its
scarcely useful algebraic complexity \cite{jup}.

\subsubsection{Toroidal modes}

As discussed above, the boundary equation for the toroidal modes in a
general viscoelastic model is obtained from the eigenvalue equation of the
elastic model:

\begin{equation}
  \beta_1(kR) = 0 \ ,\qquad   k = \sqrt{\frac{\rho}{\mu}}\,\omega
  \label{alteigen}
\end{equation}

Upon substitution of $\mu$ by $\tilde\mu$, we obtain

\begin{equation}
  \beta_1({\cal K}R) = 0\ ,\qquad
  {\cal K} = \sqrt{\frac{2\rho S'(i\Omega)}{R'(i\Omega)}}\,\Omega
\end{equation}

We know that the only solutions to the eigenvalue equation (\ref{alteigen})
are the {\it real\/} eigenfrequencies of the elastic sphere $\omega_{nl}^T$,
and therefore the allowed values for $\Omega$ are given by the implicit
relationship

\begin{equation}
  \sqrt{\frac{2\mu\,S'(i\Omega)}{R'(i\Omega)}}\,\Omega = \omega_{nl}^T
  \label{jqs}
\end{equation}

Let us now write the polynomials $S'\/$ and $R'\/$ in the form

\begin{equation}
  S'(x) = 1 + \epsilon\sum_{l=1}^{N-1} s'_lx^l\ ,\qquad
  R'(x) = 2\mu\left(1+\epsilon\sum_{l=1}^{N-1} r'_l {x}^l\right)
\end{equation}
as a suitable one to imply that internal friction effects are small, letting

\begin{equation}
   \epsilon \ll 1
\end{equation}

The quantities $r'_l\omega^l\/$ and $s'_l\omega^l\/$ are thus zero order
in $\epsilon$ and dimensionless, $\omega$ being a toroidal eigenvalue of
the elastic case. We then introduce an expansion for $\Omega$ in the
small parameter $\epsilon$, whose zeroth order term corresponds to a given
toroidal eigenfrequency $\omega\/$ of the elastic solid:

\begin{equation}
  \Omega = \omega + \epsilon\Omega_1
\end{equation}

Under the above conditions, we have

\begin{equation}
  \frac{2S'(i\Omega)}{\mu R'(i\Omega)} = 1 + 2i\epsilon\sum_{l=0}^{N-1}
   t_l\omega^l\ ,\qquad  t_l = i^{l-1}(s'_l-r'_l)/2
\end{equation}
and the value of $\Omega_1$ follows when introducing the above expansion
into equation (\ref{jqs}), yielding

\begin{equation}
   \Omega_1 = \sum_{l=1}^{N-1} t_l\omega^{l+1}
\end{equation}

In terms of $\Omega_1$, the quality factor reads

\begin{equation}
  Q = -\frac{\omega}{\epsilon}\,\frac{1}{\mbox{Im}[\Omega_1]}
\end{equation}

where Im[$\cdot$] denotes the imaginary part of its argument. Thus we
observe that, as regards toroidal modes, using a general differential
model gives us a {\it polynomial\/} in $\omega\/$ for $1/Q\/$, with no
independent term, so that constant $Q\/$ is not allowed by these models.
The polynomial only contains {\it odd\/} powers of the unperturbed
frequency $\omega$. In general, whenever $t_l\/$\,$\neq$\,0 for even $l\/$,
the real part of $\Omega_1$ will not vanish, and the angular frequency of
the periodic component of the quasinormal modes shall undergo first order
corrections. Hence, in order to preserve the elastic spectrum to first
order, our model must satisfy the conditions

\begin{equation}
   t_l = 0 \qquad \mbox{($l\/$ even)}
\end{equation}

Provided the preceeding equation holds, the corrections to ${\cal K}$ will
be purely imaginary, and therfore the modulus of the spatial part of
the modes will remain unaltered, the only effect of viscosity being the
addition of a point dependent phase in the fashion of equation (\ref{qnmM})
\footnote{
This correction will only appear provided that $N\/$\,$\geq$\,3. This is why
it was absent in the toroidal families of the previously discussed models.}.

The calculation for the spheroidal quasinormal modes can be performed along
the same lines but, as we have seen, the algebra is considerably more
involved already in the simplest models. It does naturally become more
cumbersome as the order of the model increases, so we omit a detailed
discussion of its technicalities here.

\section{Conclusions}

In this paper we have addressed the problem of whether it is possible to
systematically characterise the linewidths of the oscillation eigenmodes
of a given spherical GW detector. To this end we have considered various
phenomenological models, selected from the specialised literature on the
subject, and solved the equations of motion in the case of our interest.
Different models are seen to predict different  frequency dependences of
the quality factors for the lower modes, which are the ones we have paid
attention to, and the ones relevant for GW detection purposes. For example,
in a Kelvin-Voigt solid the $Q\/$ of a given mode appears to be inversely
proportional to its frequency, while in a Maxwell solid it is directly
proportional to it ---though significant fluctuations around these
behaviours also show in both cases.

It is not presupposed that a particular model applies to a particular
elastic material, or class of materials. Our analysis should rather help to
understand, {\it a posteriori\/} of experimental spectral measurements,
whether the solid at hand belongs in this or that phenomenological category.
This will contribute one more criterion in the selection of the most suitable
alloy the upcoming spherical GW detectors will be made with, as different
viscoelastic models (Kelvin-Voigt, Maxwell,\ldots) are associated with other
relevant properties of the solid, such as e.g.\ their response to static load
or suspension system. These are of course of the utmost practical importance
for an earth-based GW observatory.

\acknowledgements{
We thank the Spanish Ministry of Education for financial support through
contract number PB96-0384. We also acknowledge collaboration from the
Institut d'Estudis Catalans.}

\appendix

\section{}\label{aA}

A well-behaved three-dimensional vector field ${\bf s}({\bf x})$ can be
expressed as the sum of an irrotational, ${\bf s}_l({\bf x})$, and a
divergence free, ${\bf s}_t({\bf x})$, vector fields, respectively
called the {\it longitudinal\/} and {\it transverse\/} components of
${\bf s}({\bf x})$ \cite{ll}:

\begin{equation}
  {\bf s}({\bf x}) = {\bf s}_l({\bf x}) + {\bf s}_t({\bf x})\ ,
  \qquad \nabla{\bf \cdot}{\bf s}_t=\nabla\times{\bf s}_l = 0  \label{A1}
\end{equation}

We now replace this decomposition into equation (\ref{eqmovKV2}) to find

\begin{equation}
\rho\ddot T(t) ({\bf s}_t+{\bf s}_l) =
\left[(\lambda+\mu)T(t)+(\lambda'+\mu')\dot T(t)\right]
\nabla\left(\nabla\!\cdot\!{\bf s}_l\right)+\left[\mu\,T(t)+\mu'
\,\dot T(t)\right]\,\nabla^2({\bf s}_t+{\bf s}_l) \label{A2}
\end{equation}

Taking the rotational of this equation,

\begin{equation}
  \nabla\times\left[\rho\ddot T\,{\bf s}_t-(\mu\,T+\mu'\,\dot T)
  \,\nabla^2{\bf s}_t\right] = 0   \label{A3}
\end{equation} 

The vector between square brackets is thus divergence--free and irrotational,
so it vanishes. We have therefore:

\begin{equation}
  \nabla^2{\bf s}_t=\left\{\frac{\rho\,\ddot T}{\mu\,T+\mu'\,\dot T}\right\}
  \,{\bf s}_t   \label{A4}
\end{equation}

Since the left hand side of the above equation does not depend on time, the
term between braces in the right hand side must equal a (complex) constant,
say $-{\cal K}^2$. Thus,

\begin{mathletters}
 \label{A5}
  \begin{eqnarray}
    & \nabla^2{\bf s}_t+{\cal K}^2\,{\bf s}_t = 0 &  \label{A5a} \\
    & \mu\,T+\mu'\,\dot T+{\cal K}^{-2}\ddot T=0 &  \label{A5b}
  \end{eqnarray}
\end{mathletters}

An analogous procedure, after taking the divergence of equation (\ref{A2}),
gives us the corresponding formul\ae\/ for the longitudinal part:

\begin{mathletters}
 \label{A6}
  \begin{eqnarray}
    & \nabla^2{\bf s}_l+{\cal Q}^2\,{\bf s}_l = 0 &  \label{A6a} \\
    & (\lambda+2\mu)\,T+(\lambda'+2\mu')\,\dot T+{\cal Q}^{-2}\ddot T = 0 &
    \label{A6b},
  \end{eqnarray}
\end{mathletters}

where ${\cal Q}^2$ stands for another complex separation constant.

\section{}\label{aB}

We describe in this appendix the algebraic operations which lead to the
solution to the eigenvalue problem in a viscoelastic sphere. Equations
(\ref{eqmov}) ought to be solved, subject to the boundary conditions
(\ref{I.4}). The latter can be cast in explicit vector form:

\begin{equation}
  (\lambda+\gamma\lambda')\left[\nabla\!\cdot\!{\bf s}({\bf x})\right]
  {\bf n} + 2(\mu+\gamma\mu')\,({\bf n}\!\cdot\!\nabla){\bf s}({\bf x}) +
  2(\mu+\gamma\mu')\,{\bf n}\!\times\!\left[\nabla\!\times\!{\bf s}({\bf x})
  \right] = 0     \label{BCondKV}
\end{equation}

The irrotational and divergence free components, ${\bf s}_t({\bf x})$ and
${\bf s}_l({\bf x})$, can be expressed by means of auxiliary functions
$\phi({\bf x})$ and $\psi({\bf x})$:

\begin{equation}
  {\bf s}_l({\bf x}) = {\cal Q}^{-1}C_0\,\nabla\,\phi({\bf x})
  \qquad {\rm and} \qquad
  {\bf s}_t({\bf x}) = i{\cal K}^{-1}C_1\,\nabla\!\times\!{\bf L}\psi({\bf x})
  + iC_2\,{\bf L}\psi({\bf x})    \label{totKV}
\end{equation}

where $C_0$, $C_1\/$ and $C_2\/$ are (so far) undetermined integration
constants, and {\bf L}\,$\equiv$\,$-i${\bf x}$\times$$\nabla$ is the
``angular momentum'' operator. Upon substitution of (\ref{totKV}) into
(\ref{eqmov}) it is readily seen that the functions $\phi\/$ and $\psi\/$
are themselves also solutions to corresponding Helmholtz equations:

\begin{equation}
   \nabla^2\phi({\bf x})+{\cal Q}^2\phi({\bf x}) = 0 \qquad {\rm and} \qquad
   \nabla^2\psi({\bf x})+{\cal K}^2\psi({\bf x}) = 0    \label{pf}
\end{equation}

They have therefore the general form, using spherical coordinates
($r$,$\theta$,$\varphi$) for the vector {\bf x},

\begin{equation}
  \phi({\bf x}) = j_l({\cal Q}r)\,Y_{lm}(\theta,\varphi)
  \qquad {\rm and} \qquad
  \psi({\bf x}) = j_l({\cal K}r)\,Y_{lm}(\theta,\varphi)
  \label{pfsol}
\end{equation}

where $j_l\/$ are {\it spherical\/} Bessel functions of the first kind and
$Y_{lm}\/$ are spherical harmonics. The solutions (\ref{pfsol}) are those
possessing regularity properties in the whole interior and boundary of the
solid. We thus have

\begin{mathletters}
 \label{sKV}
  \begin{eqnarray}
   \nabla\phi & = & \frac{d\,j_l({\cal Q}r)}{dr}\,Y_{lm}(\theta,\varphi)\,
    {\bf n} - \frac{j_l({\cal Q}r)}{r}
    \,i{\bf n}\times{\bf L}Y_{lm}(\theta,\varphi)    \label{sKV.1} \\
   \nabla\!\times\!{\bf L}\psi & = &
    -l(l+1)\,\frac{j_l({\cal K}r)}{r}\,Y_{lm}(\theta,\varphi)\,{\bf n} +
    \left[\frac{j_l({\cal K}r)}{r} + \frac{d}{dr}j_l({\cal K}r)\right]
    \,i{\bf n}\times{\bf L}Y_{lm}(\theta,\varphi)    \label{sKV.2} \\
   {\bf L}\psi & = & j_l({\cal K}r)\,i{\bf L}Y_{lm}(\theta,\varphi)
    \label{sKV.3}
\end{eqnarray}
\end{mathletters}

These expressions ought to be substituted now into (\ref{totKV}), and then
into (\ref{BCondKV}) ---recall that {\bf s}\,=\,${\bf s}_t\/$\,+\,
${\bf s}_l\/$. It is found thet these are equivalent to the following
{\it homogeneous linear system\/}

\begin{equation}
\left(
\begin{array}{ccc}
\beta_4\left({\cal Q}R,\frac{\lambda+\gamma\lambda'}{\mu+\gamma\mu'}\right)
& -l(l+1)\frac{{\cal K}}{{\cal Q}}\beta_1({\cal K}R) & 0 \\
-\beta_1({\cal Q}R) & \frac{{\cal K}}{{\cal Q}}\,\beta_3({\cal K}R) & 0 \\
0 & 0 & -\frac{{\cal K}}{{\cal Q}}\,{\cal K}R\,\beta_1({\cal K}R)
\end{array}\right)\left(
\begin{array}{c}
C_0 \\ C_1 \\ C_{2}
\end{array}\right)=0 \label{bcatlastKV}
\end{equation}
where

\begin{mathletters}
 \label{betas}
  \begin{eqnarray}
   \beta_1(z) & \equiv & \frac{d}{dz}\left[\frac{j_l(z)}{z}\right] \\
   \beta_2(z) & \equiv & \frac{d^2}{dz^2}\left[j_l(z)\right] \\
   \beta_3(z) & \equiv & \mbox{\large $\frac{1}{2}$}\,\beta_2(z) +
     \left\{\mbox{\large $\frac{l(l+1)}{2}$}-1\right\}\,\beta_0(z) \\
   \beta_4(z,A) & \equiv & \beta_2(z)-\frac{A}{2}j_l(z)
  \end{eqnarray}
\end{mathletters}

The system (\ref{bcatlastKV}) is to be satisfied by the constants
$C_0$, $C_1\/$ and $C_2\/$, but has no meaningful solution unless the
system matrix is {\it singular\/}, i.e., if its determinant vanishes.
Therefore

\begin{equation}
 \beta_1({\cal K}R)\,\det\left(
  \begin{array}{cc}
  \beta_4\left({\cal Q}R,\frac{\lambda+\gamma\lambda'}{\mu+\gamma\mu'}\right)
   & -l(l+1)\frac{{\cal K}}{{\cal Q}}\beta_1({\cal K}R) \\
   -\beta_1({\cal Q}R) & \frac{{\cal K}}{{\cal Q}}\beta_3({\cal K}R)
  \end{array}\right) = 0        \label{eigen}
\end{equation}

This is an equation for the parameter $\gamma\/$, on which ${\cal K}$ and
${\cal Q}$ depend through equations (\ref{KQ}). Clearly, there are {\it two\/}
families of solutions, or {\it eigenmodes\/}, to (\ref{eigen}) associated to
the vanishing of either of the two factors in its lhs, i.e.,
$\beta_1({\cal K}R)$ {\it or\/} the determinant of the displayed 2$\times$2
matrix. They are called {\it toroidal\/} and {\it spheroidal\/} solutions,
respectively.

\end{document}